\documentclass[twoside,twocolumn,letterpaper]{article}

\usepackage{a4wide}
\usepackage{abstract}
\usepackage{amsmath}
\usepackage[british]{babel}
\usepackage{colortbl}
\usepackage{fancyhdr}
\usepackage{fancyvrb}
\usepackage[top = 1in, left =1in, right = 1in, bottom = .75in]{geometry}
\usepackage{graphicx}
\usepackage{hyperref}
\usepackage{lastpage}
\usepackage{relsize}
\usepackage{subfig}
\usepackage[usenames,dvipsnames,svgnames,table]{xcolor}

\graphicspath{{../../Output/Paper1/}}

\title{Reproducible probe-level analysis of the Affymetrix Exon 1.0 ST array with R/Bioconductor}
\author{Maria Rodrigo-Domingo, Rasmus Waagepetersen, Julie St\o ve B\o dker, Steffen Falgreen, \\ Malene Krag Kjeldsen, Hans Erik Johnsen, Karen Dybk\ae r and Martin B\o gsted}
\date{}

\begin{document}
\maketitle 

\begin{abstract}
Alternative splicing is the post-transcriptional process by which a single gene can produce multiple transcripts and thereby protein isoforms. The presence of different transcripts of a gene across samples can be analysed by whole-transcriptome microarrays. Reproducing results from published microarray data represents a challenge due to the vast amounts of data and the large variety of pre-processing and filtering steps employed before the actual analysis is carried out. To ensure a firm basis for methodological development where results with new methods are compared with previous results it is crucial to ensure that all analyses are completely reproducible for other researchers.

We here give a detailed workflow on how to perform reproducible analysis of the GeneChip\textregistered\ Human Exon 1.0 ST Array at probe and probeset level solely in \texttt{R}/Bioconductor, choosing packages based on their simplicity of use. To exemplify the use of the proposed workflow we analyse differential splicing and differential gene expression in a publicly available dataset using various statistical methods.

We believe this study will provide other researchers with an easy way of accessing gene expression data at different annotation levels and with the sufficient details needed for developing their own tools for reproducible analysis of the  GeneChip\textregistered\ Human Exon 1.0 ST Array.\\
\textbf{Contact:} \href{maria@math.aau.dk}{maria@math.aau.dk}
\end{abstract}

\section{Introduction}

In the field of microarrays it has traditionally been difficult to compare new methods to methods in published papers as the many methods available for pre-processing, summarizing and filtering make it almost impossible to work with the exact same data, even when the raw data is made available. That is why we consider reproducible research fundamental as it will facilitate easy 1) revision of papers, 2) access to data and results, 3) communication to other researchers and 4) comparison between different methods. Reproducible research is gaining relevance among the scientific community as shown by the number of papers published on the subject during the last years \cite{Baggerly2010, Baggerly2009, Coombes2007}. Ioannidis \textit{et al.} showed that the results of only 2 out of 18 published microarray gene-expression analyses were completely reproducible \cite{Ioannidis2009}. This is why some authors demand that documentation and annotation, database accessions and URL links and even scripts with instructions are made publicly available \cite{Baggerly2010}. Journals like Biostatistics have even appointed an Associate Editor for reproducible research, but still treat it as a ``desirable goal" rather than a requirement \cite{Peng2009}. Setting up a framework for reproducible research necessarily implies working with free and open-source software as for example R/Bioconductor \cite{R,Bioconductor}. Additionally, using Sweave \cite{Sweave} (a tool for embedding \texttt{R} code in \LaTeX documents \cite{Latex}), enables automatic reports that can be updated with output from the analysis.

The main tool in this paper will be the Bioconductor package \texttt{aroma.affymetrix} \cite{aroma} that can analyse all Affymetrix chip types with a (.CDF) Chip Definition File. The number of arrays (samples) that can  be simultaneously analysed by \texttt{aroma.affymetrix} is virtually unlimited as the system requirements are just 1GB RAM, for any operating system \cite{aromaProject}. This package is freely available and can easily be installed into \texttt{R}. The aroma.affymetrix website \url{www.aroma-project.org} is conceived as reference for all the possible microarrays that can be analysed with \texttt{aroma.affymetrix}, and does not focus specifically on the analysis of the GeneChip\textregistered\ Human Exon 1.0 ST Array (or exon array in short). Portable scripts for a fast and basic analysis can be obtained on request to \texttt{aroma.affymetrix}'s authors.

The analysis of exon array data in \texttt{R}/Bioconductor is not yet standard. There are several packages available, and it can be a tremendous effort for a newcomer to maneuver between them, and to overcome the numerous challenges associated with these packages. This paper aims to make this task easier and to provide a quick reference guide to \texttt{aroma.affymetrix}'s documentation. We also explain how to extract data for different statistical analyses and propose a method for gene annotation and for gene profile visualization. For this last step, we use the packages \texttt{biomaRt} and \texttt{GenomeGraphs} to annotate and visualize the transcripts in a genomic context.
 
The analysis workflow presented in this paper is carried out solely in \texttt{R}/Bioconductor, and the paper is available as a Sweave (.Snw) document that will allow the reader to reproduce our exact results. The .Snw document can also be converted into an \texttt{R} script and executed. The workflow starts by reading in the data, followed by background correction and quantile normalization. We then explain how to obtain transcript cluster -, probeset -, and probe-level estimates. Afterwards, different methods for the statistical analysis of differential splicing or differential gene expression are reviewed. Finally, we make a suggestion on how to annotate transcript clusters to genes in the lists obtained from the statistical analyses, and how to plot the data including genomic information. To exemplify the use of the workflow, an example dataset \cite{Gardina2006} is analysed along the way.

\section{Background on alternative splicing}
Splicing is the post-transcriptional process that generates mature eukaryotic mRNAs from pre-mRNAs by removing the non-coding intronic regions and joining together the exonic coding regions. For many genes, two or more splicing events take place during maturation of mRNA molecules resulting in a corresponding number of alternatively spliced mRNAs. These mature mRNAs translate into protein isoforms differing in their amino acid sequence and ultimately in their biochemical and biological properties \cite{Black2003, Hallegger2010}. Alternative splicing is one of the main tools for generating RNA diversity, contributing to the diverse repertoire of transcripts and proteins \cite{Hallegger2010, Licatalosi2010}. It is known that 92-94\% of multi-exon human genes are alternatively spliced and that 85\% of those have a minor isoform frequency of at least 15\% \cite{Pan2008, Wang2008}. In our case we will focus on the detection of differential splicing between groups, as for instance tissue types, or healthy vs. diseased samples. 

The exon array was presented in October 2005 as a tool for the analysis and profiling of whole-transcriptome expression \cite{Clark2007, Suzuki2011, Thorsen2011}. To interrogate each potential exon with at least one probeset, the exon array contains about 5.6 million probes grouped into more than 1.4 million probesets (most probesets consisting of 4 probes), which are further grouped into 1.1 million exon clusters, or collections of overlapping exons. Finally, exon clusters are grouped into over 300.000 transcript clusters to describe their relationship, as for example shared splice sites or overlapping exonic sequences. Each gene is covered, on average, by 40 probes interrogating regions located along the entire gene \cite{AffymetrixWholeTranscriptExpression}. This probe positioning aims at providing better estimates of gene expression levels than previous arrays, and allows for the study of differential splicing \cite{Lockstone2011} and differential gene expression based on summarized exon expression.  

The exon array has three levels of annotation for the interrogated transcript clusters: \textit{core, extended} and \textit{full} \cite{AffymetrixAlternativeSplicing}. Core transcript clusters are supported by the most reliable evidence such as RefSeq transcripts and full-length mRNAs \cite{Pruitt2007} and a core transcript cluster is roughly a gene \cite{AffymetrixExonDesign}; the extended level contains the core transcript clusters plus cDNA-based annotations \cite{Benson2012} and the full level contains the two previous levels plus \textit{ab-initio}, or algorithmic, gene predictions \cite{Burge1997}. It is worth noting that \texttt{aroma.affymetrix} enables the analysis at the three levels of annotation mentioned above, and also that it provides intensity estimates for probes, probesets and transcript clusters, allowing for a variety of options for the analysis.

\section{Workflow}
\begin{figure*}[!tpb] 
\centerline{\includegraphics[width=1\textwidth]{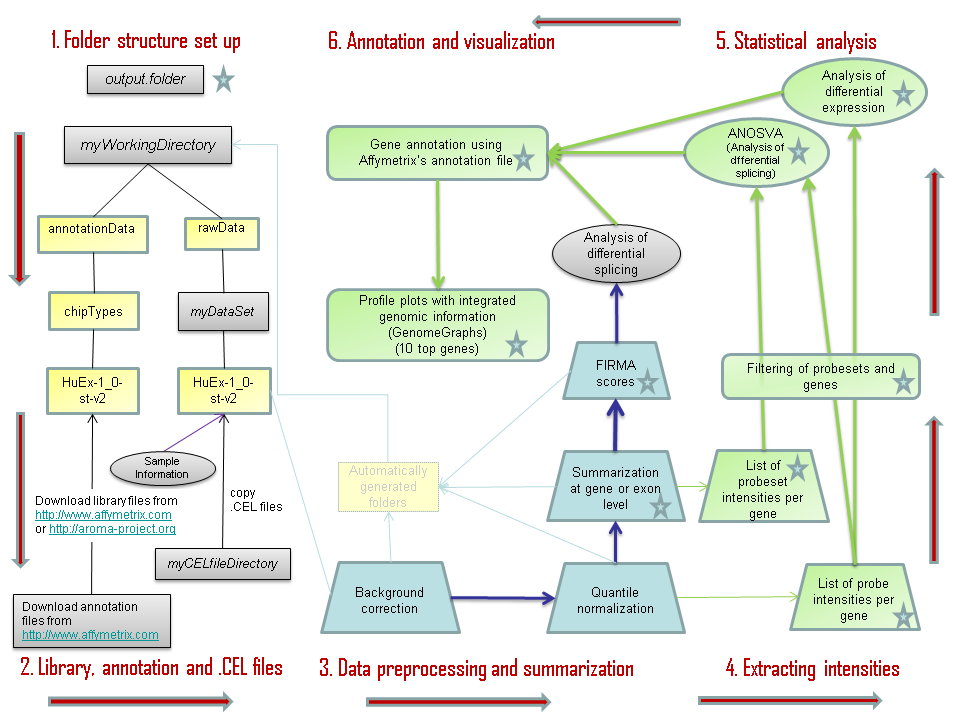}}
		\caption{\label{fig:flowChart} Flowchart for an analysis with \texttt{aroma.affymetrix}, read counter-clockwise starting in upper left corner: 1. and 2. folder structure set-up including library, annotation and .CEL files, 3. data pre-processing and summarization, 4. extraction of intensities at transcript cluster, probeset and probe level, including filtering recommended by Affymetrix, 5. statistical analysis of differentially expressed or spliced transcript clusters and 6. annotation and visualization of transcript cluster profiles. Blue boxes represent parts of the analysis implemented in \texttt{aroma.affymetrix}, yellow and green boxes are part of the code provided in this paper, and purple boxes represent user-input needed. Output produced at several steps is saved in user-chosen "\texttt{output.folder}" and represented by a star shape in the workflow. A number of folders are automatically generated by \texttt{aroma.affymetrix} - represented by a faded yellow rectangle in 3. -, our workflow does not make use of the contents of such folders.}
\end{figure*}

Our workflow for the analysis of exon array data starts setting up the required folder structure for \texttt{aroma.affymetrix}. The data is then preprocessed and summarised at transcript cluster and/or probeset level. Next, transcript clusters are analysed with several statistical models to detect differential expression or splicing and the transcripts of interest are annotated  and visualised at the end, see Figure \ref{fig:flowChart}. In the code, places where user input is needed are marked by ``***", and places where the user can choose whether to modify parameters are marked by ``**".

To exemplify the use of the tutorial we have used Affymetrix's colon cancer data set \cite{Gardina2006}, consisting of a collection of paired samples of colon tumour tissue and adjacent normal tissue from 10 patients and available at \url{http://www.affymetrix.com/support/technical/sample_data/exon_array_data.affx}. According to Affymetrix's website, the RNA samples are from a commercial source. This dataset has been used in a number of papers to evaluate the performance of different analysis methods \cite{Purdom2008, Zheng2009, Affymetrix2005, Turro2010} and a number of genes have been validated to present differential splicing or not \cite{Gardina2006}. The analysis was done in \texttt{R} version 2.15.1 (32 bit).

Start by installing and loading \texttt{aroma.affymetrix} in \texttt{R} and loading the other libraries required: 

\begin{Verbatim}[fontfamily=courier, fontshape=it, fontsize = \relsize{-1.35}]
> source("http://aroma-project.org/hbLite.R")
> hbInstall("aroma.affymetrix")
> require(aroma.affymetrix)
> require(biomaRt)
> require(GenomeGraphs)
\end{Verbatim}

\subsection{Setting up the structure and files for the analysis workflow} \label{subsec:folderAndFiles}
This section corresponds to steps 1. and 2. in Figure \ref{fig:flowChart}. The first step is to create the folder structure: under a main folder of our choice - ``\texttt{myworkingDirectory}" - we will create the ``\texttt{rawData}" and ``\texttt{annotationData}" folders, which will be common to all \texttt{aroma.affymetrix} projects. Inside ``\texttt{annotationData}", the subfolder ``\texttt{chipTypes}" will contain one subfolder per chip type, with the exact name of the .CDF file provided by Affymetrix, ``\texttt{HuEx-1\_0-st-v2}" in our case. Inside this folder we will save any library and annotation files that might be needed. Besides, the ``\texttt{myDataSet}" folder will be created under ``\texttt{rawData}" to store .CEL files. These files are the output of a microarray experiment and contain the result of the intensity calculations per probe or pixel. Note that the microarray experiment produces one .CEL file per array and that one array analyses one sample. Affymetrix sometimes refer to their microarrays as ``chips". Note also that we need one ``\texttt{myDataSet}" folder per experiment and that ``\texttt{myDataSet}" will be added as a tag at the end of the \texttt{aroma.affymetrix} output.

\begin{Verbatim}[fontfamily=courier, fontshape=it, fontsize = \relsize{-1.35}]
> #*** user-defined working directory
> wd <- "myWorkingDirectory"
> #*** user-defined data set name
> ds <- "myDataSet"
\end{Verbatim}

In the second step we save our library (Chip Definition File (.CDF) in our case) and .CEL files in the corresponding folders. Affymetrix's unsupported .CDF files can be downloaded from \begin{small}\url{http://www.affymetrix.com/Auth/support/downloads/library_files/HuEx-1_0-st-v2.cdf.zip}\end{small}, note that registration is needed. For the exon array, Elizabeth Purdom has created a number of binary .CDF files based on Affymetrix's text .CDF file \cite{aromaProject} that are faster to query and more memory efficient. Such binary .CDF files for core, extended and full sets of probesets can be downloaded from \url{http://aroma-project.org/node/122}. In the example below we use the custom aroma file for core transcript clusters, which might be updated in the future. Our original .CEL files will be copied from the user-specified ``\texttt{myCELfileDirectory}" into the exon ``\texttt{rawData}" subfolder (the code is part of the .Snw version of this paper). The desired output folder specified in ``\texttt{output.folder}" should exist in advance.

\begin{Verbatim}[fontfamily=courier, fontshape=it, fontsize = \relsize{-1.35}]
> #** download user-defined library file
> library.file <- 
+   paste(annotation.data.exon, 
+         "HuEx-1_0-st-v2,coreR3,A20071112,EP.cdf",
+         sep = "/")
> download.address <- 
+      "http://bcgc.lbl.gov/cdfFiles/"
> file <- paste("HuEx-1_0-st-v2,A20071112,EP", 
+             "HuEx-1_0-st-v2,coreR3,A20071112,EP.cdf",
+             sep = "/")
> custom.cdf <- 
+      paste(download.address, file, sep = "")
> download.file(url = custom.cdf, 
+               destfile = library.file, 
+               mode = "wb", quiet = FALSE)
> #*** user-defined directory containing .CEL files
> cel.directory  <- "myCELfileDirectory" 
> #*** user-defined output folder
> output.folder <- "output.folder"
\end{Verbatim}

Besides, the sample information should be saved in a tab separated file with column names \texttt{celFile}, \texttt{replicate}, and \texttt{treatment} containing .CEL file name (without .CEL), replicate identifier and treatment name, respectively. This file should be called "SampleInformation.txt" and it will be copied from the user-specified directory into the \texttt{"\textbackslash rawData\textbackslash myDataSet \textbackslash HuEx-1\_0-st-v2"} folder (which will also contain the .CEL files) after it has been created. The sample information file for the colon cancer example is attached as an additional file.

\begin{Verbatim}[fontfamily=courier, fontshape=it, fontsize = \relsize{-1.35}]
> sample.info <- 
+     read.table(file = paste(raw.data.exon, 
+                "SampleInformation.txt", sep = "/"),
+                sep = "\t", header = TRUE)
\end{Verbatim}

Finally, NetAffx transcript clusters' and probesets' annotation files should be saved in ``\texttt{annotationData/chipTypes/}'' ``\texttt{HuEx-1\_0-st-v2}''. We have used release 32, which was most up to date at the time of writing and we downloaded files  ``HuEx-1\_0-st-v2.na32.hg19.transcript.csv.zip'' and ``HuEx-1\_0-st-v2.na32.hg19.probeset.csv.zip'' from \begin{footnotesize}\url{http://www.affymetrix.com/estore/browse/products.jsp?productId=131452&categoryId=35676&productName=GeneChip-Human-Exon-ST-Array#1_3}\end{footnotesize}, \textit{Technical Documentation} tab, under \textit{NetAffx Annotation Files}. The extracted .csv files should be converted into .Rdata files for querying them faster in the future. Note that the number of lines to skip might differ for future annotation files. 

\begin{Verbatim}[fontfamily=courier, fontshape=it, fontsize = \relsize{-1.35}]
> transcript.clusters.NetAffx.32 <- 
+     read.csv(file = paste(annotation.data.exon,
+         "HuEx-1_0-st-v2.na32.hg19.transcript.csv", 
+         sep = "/"), skip=24)
> probesets.NetAffx.32 <- 
+ read.csv(file = paste(annotation.data.exon,
+          "HuEx-1_0-st-v2.na32.hg19.probeset.csv", 
+          sep = "/"), skip=23) 
\end{Verbatim}

\subsection{Data pre-processing and summarization to probeset/transcript cluster level} \label{subsec:pre-processing}
After defining chip type and dataset, background correction and quantile normalization are carried out as shown in Figure \ref{fig:flowChart}, step 3. In these pre-processing steps, it is possible to use either Affymetrix's original .CDF file, the .CDF files provided by the aroma project (the file for core transcript clusters in our example), or a .CDF file created by the user. The summarization step, however, must be done using one of the custom .CDF files available at the \texttt{aroma.affymetrix} project website.

Background correction as defined by Irizarry \cite{Irizarry2003Exploration} and quantile normalization are performed by the \texttt{RmaBackgroundCorrection()} and \texttt{QuantileNormalization()} functions, respectively. The raw, background corrected and quantile normalized probe intensities can be visualized using the \texttt{plotDensity()} function applied to the corresponding object. Summarization is done with the \texttt{ExonRmaPlm} function \cite{Irizarry2003Summaries}. The parameter \texttt{mergeGroups} determines whether to summarize at transcript level (\texttt{TRUE}) or probeset level (\texttt{FALSE}). All the functions described automatically create subfolders such as ``plmData" or ``probeData" inside ``\texttt{myWorkingDirectory}". A more detailed version of this code with interesting comments about the choice of .CDF and possibilities for quality control is available at \url{http://www.aroma-project.org/vignettes/FIRMA-HumanExonArrayAnalysis}. 

\begin{Verbatim}[fontfamily=courier, fontshape=it, fontsize = \relsize{-1.35}]
> chipType <- "HuEx-1_0-st-v2" 
> #** user-defined .CDF file: change tags parameter
> cdf <- AffymetrixCdfFile$byChipType(chipType, 
+                    tags = "coreR3,A20071112,EP")
> cs <- AffymetrixCelSet$byName(ds, cdf = cdf) 
> 
> # background correction
> bc <- RmaBackgroundCorrection(cs) 
> csBC <- process(bc, verbose = verbose) 
> 
> # quantile normalization
> qn <- QuantileNormalization(csBC, 
+                typesToUpdate = "pm")
> csN <- process(qn, verbose = verbose)
> 
> # summarization
> # transcript cluster level
> plmTr  <- ExonRmaPlm(csN, mergeGroups = TRUE, 
+                tag = "coreProbesetsGeneExpression")
> # probeset/exon level
> plmEx  <- ExonRmaPlm(csN, mergeGroups = FALSE, 
+                tag = "coreProbesetsExonExpression")			
\end{Verbatim}

\subsection{Extraction of intensity estimates} \label{subsec:extractingIntensities}
The \texttt{aroma.affymetrix} documentation focuses on analyses at the probeset and transcript cluster levels. The respective intensities are obtained by applying the function \texttt{getChipEffectSet()} to the transcript or probeset \texttt{plm} objects (\texttt{plmTr} and \texttt{plmEx}, respectively) and then extracting the corresponding dataframes. However, it is also possible to extract the background corrected and quantile normalized intensities of all probes using the function \texttt{getUnitIntensities}. While \texttt{plmTr} is suitable for the FIRMA analysis (\ref{subsec:statistics}), \texttt{plmEx} is well suited for probeset-level analysis. For the linear model analysis described in Section \ref{subsec:statistics} we have created one list of dataframes containing probe intensities per transcript cluster, and another list of dataframes containing probeset intensities per cluster (code included in .Snw).

\begin{Verbatim}[fontfamily=courier, fontshape=it, fontsize = \relsize{-1.35}]
> # extract a matrix of gene intensities
> cesTr <- getChipEffectSet(plmTr)
> trFit <- extractDataFrame(cesTr, addNames=TRUE) 
> # extract a matrix of probeset intensities
> cesEx <- getChipEffectSet(plmEx)
> exFit <- extractDataFrame(cesEx, addNames=TRUE) 
> # extract a list of probe intensities per gene
> unitIntensities <- 
             readUnits(csN, verbose=verbose) 
\end{Verbatim}

The high number of transcript clusters analysed in combination with the usually small number of chips tends to cause a high number of false positives \cite{AffymetrixAlternativeSplicing}. In order to reduce the number of false positives, Affymetrix recommends to perform detection above background (DABG) \cite{DABG} on the dataset prior to the analysis \cite{AffymetrixAlternativeSplicing}. The DABG procedure is not implemented in \texttt{aroma.affymetrix}, so we decided to follow the procedure described in \cite{Purdom2008} and use 3 as a threshold for the probeset intensity, so that probesets with a $\log_2$ intensity below 3 will be marked as absent. Except for this change, we followed the guidelines proposed in \cite{AffymetrixAlternativeSplicing} to remove absent transcript clusters and probesets, where neither probesets that are absent in more than half of the samples of a group nor transcript clusters with more than half of the probesets absent are analysed.

Besides this filtering based on expression levels, another filtering that removes probesets presenting cross-hybridization is also advisable \cite{AffymetrixAlternativeSplicing}. Cross-hybridising probesets are identified in file "HuEx-1\_0-st-v2.na32.hg19.probeset.csv" and removed. Affymetrix recommends to filter them out after the analysis, but we have decided not to include them in the analysis in order to narrow down the number of probesets/transcript clusters to investigate.

The filtering procedure is part of the .Snw file. In our example, where we analysed only core probesets, 136233 probesets out of 284258 were deemed present by our filter, and the number of transcript clusters to analyse (present in both samples) was reduced from 18708 to 8401.

\subsection{Statistical analysis} \label{subsec:statistics}
In this section we give an overview of model-based statistical methods available for the analysis of differential splicing and suggest a method for the analysis of differential gene expression.

\subsubsection{Differential splicing}\label{subsubsec:DS}
The analyses carried out in this paper are model-based approaches, and the analysis of differential splicing is done genewise. The models we use are extensions of the linear model by Li and Wong \cite{Li2001}
\begin{equation}
y_{pt} = \alpha_p + \beta_t + \varepsilon_{pt}, \label{eq:Li2001}
\end{equation}
where $y_{pt}$ is the intensity measure of probe $p$ for treatment $t$, $\alpha_p$ is a probe affinity term, $\beta_t$ is the gene-level estimate for treatment $t$, and $\varepsilon_{pt}$ is the error term. 

\textbf{ANOSVA} (Analysis of Splicing Variation) was presented by Cline \textit{et al.} \cite{Cline2005} and is a two-way ANOVA model with probeset and treatment as factors: 

\begin{equation}
y_{pet} = \mu + \alpha_e + \beta_{t} + \gamma_{et} + \varepsilon_{pet}, \label{eq:Cline2005}
\end{equation}
where $y_{pet}$ is the intensity measure of probe $p$ in probeset $e$ and treatment $t$, the overall mean $\mu$ is the baseline level of all probes in all experiments, and $\alpha_e$ and $\beta_t$  are the probeset and treatment effects. The interaction term $\gamma_{et}$ tells whether the effect of the probeset depends on the treatment and is therefore key to the detection of differential splicing. 

The model in \eqref{eq:Cline2005} can be extended to include random effects associated to replications from the same individual, $r$:
\begin{equation}
y_{petr} = \mu + \alpha_e + \beta_{t} + \gamma_{et} + I_r + C_{tr} + \varepsilon_{petr}, \label{eq:meANOSVA}
\end{equation}
where $I_r$ is the random effect of each individual $r$ and $C_{tr}$ is the random chip effect. The error terms $\varepsilon_{petr}$ are i.i.d. $\text{N}(0, \sigma^2)$-distributed.

Under the null-hypothesis of no differential splicing, the $\gamma_{et}$'s will all be zero, therefore we consider the test statistic
$$t_{et}=\frac{\hat{\gamma}_{et}}{\hat{\sigma}},$$
where $\hat{\gamma}_{et}$ is the estimate for $\gamma_{et}$, and $\hat{\sigma}$ is the standard error of $\hat{\gamma}_{et}$. Large values will be critical for the null hypothesis. Under the model assumptions, $t$ will follow a $t_{N - T\cdot(n_e + R -1)}$ distribution \cite[Chap. 5]{Chambers1992}, with $N$ the total number of observations per transcript cluster, $T$ the number of treatments, $n_e$ the number of probesets in the transcript cluster and $R$ the number of individuals. For each gene, the smallest $p$-value from the above $t$-tests is regarded as a measure of confidence that the gene is differentially spliced across the experimental conditions \cite{Cline2005}. The interaction estimates and variances and thus the test statistics are contrast-dependent, so choosing a different contrast will alter the gene lists. In our analysis, we have used the \texttt{sum} contrast available in \texttt{R}, where parameter estimates are centred around zero. 

A package in \texttt{R} for fitting linear and generalized linear mixed-effects models is \texttt{lme4} \cite{lme4}. Linear mixed-effects models can also be analysed using \texttt{lm}, which is faster but requires a balanced design. In our example dataset we use \texttt{lm} to fit the model in equation (\ref{eq:meANOSVA}) to the probe level estimates obtained using \texttt{unitIntensities()} from \texttt{aroma.affymetrix} to 8075 multiexonic transcript clusters. Here, we only show the code corresponding to equation \eqref{eq:meANOSVA}, the rest of the code is part of the .Snw file.

\begin{Verbatim}[fontfamily=courier, fontshape=it, fontsize = \relsize{-1.35}]
> # ** user-defined parameters for linear model
> lm <- 
+     lm(intensity ~ probeset + treatment + 
+                  C(probeset:treatment, contr.sum) + 
+                  replicate/treatment)
> n.probesets <- length(unique(dataframe$probeset))
> main.effects <- 
+            1 + (n.probesets - 1) + 
+            (length(unique(dataframe$treatment)) - 1) 
> DS.parameters <- (n.probesets - 1)*
+           (length(unique(dataframe$treatment)) - 1)
> p.t <- 
+     min(summary(lm)$coefficients[(main.effects+1):
+       (main.effects + DS.parameters),"Pr(>|t|)"])
\end{Verbatim}

Although the vast majority of probesets contain 4 probes, transcript clusters containing probesets with less than 4 probes will give rise to an unbalanced design. Nevertheless, the $t$-distribution is almost a normal distribution for long transcript clusters so the unbalanced design does not have any practical implications. For shorter transcript clusters, however, an unbalanced design might be a problem.

The top 10 most differentially spliced genes, sorted by the minimum $p$-value of their $t$-tests, appear in Table \ref{tab:ANOSVAprobe}. The gene \textit{ZAK} was validated as differentially spliced in \cite{Gardina2006}. See Figure \ref{fig:ANOSVAprobe} for the profile plot of \textit{KLK10}, where the thick lines representing the mean intensity in each group have been plotted for easing the interpretation. Note that there is one measurement per probe in each probeset, typically 4 probes per probeset. How to obtain such plots is described in Section \ref{subsec:annotationAndVisualization} below. The genes in Figures \ref{fig:ANOSVAprobe}, \ref{fig:ANOSVAprobeset}, \ref{fig:FIRMA} and \ref{fig:DEprobe} were chosen because they span over a shorter genomic region and show a more clear picture of the relationship between probesets and exons than the other genes in the lists of top 10 genes.

A slight variation of ANOSVA is the probeset-model as implemented in Partek \cite{Partek} (note that the probe subscript $p$ has been removed):
\begin{equation}    
y_{etr} = \mu + \alpha_e + \beta_{t} + \gamma_{et} + I_r + C_{tr} + \varepsilon_{etr}. \label{eq:meExonANOSVA}
\end{equation}

For the probeset-level ANOSVA we used the probeset level estimates obtained by \texttt{affyPLM(..., mergeGroups = FALSE)}. After filtering for non-present or cross-hybridising probesets, and absent transcript clusters, we were left with 9494 transcript clusters to study. These clusters are sorted according to the minimum $p$-value of the individual $t$-test scores for differential splicing, and the top 10 genes obtained appear in Table \ref{tab:ANOSVAprobeset}. The genes \textit{MMP11, ZAK} and \textit{COMP} are in the top 10 genes for both the ANOSVA probe and the ANOSVA probeset models. Gene \textit{TGFBI} appears in Figure \ref{fig:ANOSVAprobeset}.

\textbf{FIRMA} (Finding Isoforms using Robust Multichip Analysis) was first introduced by Purdom \textit{et al.}  \cite{Purdom2008} for the exon array. In presence of differential splicing the model in \eqref{eq:Li2001} will not fit and this will show up in the residuals. The linear model used is the following:
\begin{equation}
y_{petr} = \alpha_{p} + \beta_{tr} + \varepsilon_{petr},\label{eq:FIRMA}
\end{equation}
with $\alpha_p$ the probe affinity, $\beta_{tr}$ the gene-level effect for chip $tr$ and the error terms $\varepsilon_{petr}$ are i.i.d. $\text{N}(0, \sigma^2)$-distributed. Note that in contrast to the model in \eqref{eq:meANOSVA} we do not compute an overall gene-level estimate, but a gene-level estimate per chip.

As in model \eqref{eq:Li2001}, there is no treatment/probeset interaction term, so differential splicing is analysed probesetwise, using the residuals per probeset $e$:
\begin{align} \label{eq:FIRMAresiduals}
r_{petr} & = y_{petr} - (\hat{\alpha}_p + \hat{\beta}_{tr}), \\ 
p = 1, \ldots n_e,&\ t = 1, \ldots, T, \ r = 1, \ldots, R \nonumber
\end{align}
where $\hat{\alpha}_p$ and $\hat{\beta}_{tr}$ are the estimates of $\alpha_p$ and $\beta_{tr}$. 

The median of the standardized residuals per probeset per chip is chosen as score statistic:
\begin{equation}
F_{e(tr)}=\substack{\textrm{median}\\{p=1,\ldots,n_e}}\left(\frac{r_{petr}}{\text{MAD}}\right). \label{eq:FIRMAScore}
\end{equation}
The standardization with the median absolute deviation (MAD) of the residuals per gene makes the scores comparable across transcript clusters.  

The FIRMA scores are extracted from the \texttt{plmTr} object obtained in Section \ref{subsec:pre-processing}. All probesets and transcript clusters were analysed by FIRMA, since it is part of the default \texttt{aroma.affymetrix} workflow. 

\begin{Verbatim}[fontfamily = courier, fontshape = it, fontsize = \relsize{-1.35}]
> firma <- FirmaModel(plmTr)
> fit(firma, verbose = verbose)
> fs <- getFirmaScores(firma)
> firma.scores <- extractDataFrame(fs)
\end{Verbatim}

After obtaining the FIRMA scores per probeset per sample, we proceeded as described in \cite{Purdom2008}: 1) For each probeset we took the difference of FIRMA scores for each of the 10 pairs of normal/cancer samples and 2) calculated the mean of the 10 differences per probeset. Then 3) we ranked the probesets according to their absolute mean difference: as the scores are comparable across transcript clusters, larger average differences between the normal and cancer samples will point at exons more differentially spliced between the two conditions. The resulting list was filtered to keep only probesets and transcript clusters that has passed the filter described in \ref{subsec:extractingIntensities} above, the code is part of the .Snw file. In order to get a gene list instead of a probeset list as in \cite{Purdom2008} we mapped probesets to transcript clusters and then selected the top 10 genes on the list, see Table \ref{tab:FIRMA}. The profile plot of gene \textit{LGALS4} appears in Figure \ref{fig:FIRMA}. The very high average difference for this gene is due to a FIRMA score of 830030.8 at probeset \textit{3861578} corresponding to the tumour sample of replicate 7.

Out of 14 genes investigated for differential splicing in \cite[Table 1]{Gardina2006}, 10 passed our filtering procedure and were analysed for differential splicing: \textit{ACTN1, VCL, CALD1, SLC3A2, COL6A3, CTTN, FN1, MAST2, ZAK} and \textit{FXYD6}. The gene \textit{ZAK} appears in the top 10 genes from ANOSVA probe and ANOSVA probeset, but a plot is not produced automatically by the code because \textit{ZAK} is not recognised by BioMart. Instead, we looked the gene up in PubMed obtaining the Ensembl ID: ENSG00000091436, and used this to plot the gene, see Figure \ref{fig:ZAK}. The gene \textit{COL6A3} appears among the top 100 genes for the ANOSVA probe and the ANOSVA probeset methods. Gene \textit{ACTN1} is among the top 100 genes for ANOSVA probeset and it is also the first of Gardina's genes to appear in the filtered FIRMA list, at position 154. 

\begin{table}[htb]
\centering
\footnotesize
\subfloat[][Model: ANOSVA probe. \label{tab:ANOSVAprobe}]{
\begin{tabular}{|l|r|}
\hline
Gene symbol & min $p$-value\tabularnewline
\hline\hline
MMP11&$2.98e-10$\tabularnewline
SOX9&$2.59e-09$\tabularnewline
FOXQ1&$6.38e-09$\tabularnewline
\rowcolor{gray} KLK10&$3.84e-08$\tabularnewline
SYNM&$4.53e-07$\tabularnewline
PHLDA1&$4.85e-07$\tabularnewline
\rowcolor{gray} ZAK&$4.89e-07$\tabularnewline
SNTB1&$5.33e-07$\tabularnewline
COMP&$6.85e-07$\tabularnewline
XPOT&$7.87e-07$\tabularnewline
\hline
\end{tabular}
}
\centering 
\subfloat[][Model: ANOSVA probeset. \label{tab:ANOSVAprobeset}]{
\begin{tabular}{|l|r|}
\hline
Gene symbol & min $p$-value\tabularnewline
\hline\hline
SOX4&$9.07e-15$\tabularnewline
MMP11&$9.48e-14$\tabularnewline
\rowcolor{gray} ZAK&$4.40e-13$\tabularnewline
FOXQ1&$5.98e-13$\tabularnewline
\rowcolor{gray} TGFBI&$4.83e-12$\tabularnewline
UBAP2L&$8.17e-12$\tabularnewline
COMP&$2.87e-11$\tabularnewline
SLC2A1&$4.16e-11$\tabularnewline
CDH11&$4.19e-10$\tabularnewline
CPXM1&$4.47e-10$\tabularnewline
\hline
\end{tabular}
}
\\
\centering
\subfloat[][Model: FIRMA. \label{tab:FIRMA}]{
\centering
\begin{tabular}{|l|r|}
\hline
Gene symbol & Average difference\tabularnewline
\hline\hline
\rowcolor{gray} LGALS4&$83000.0$\tabularnewline
HMGCS2&$   82.9$\tabularnewline
RPS6KA1&$   31.8$\tabularnewline
MUC13&$   28.2$\tabularnewline
RPL35&$   24.5$\tabularnewline
RPL35A&$   18.8$\tabularnewline
SLC39A14&$   18.2$\tabularnewline
TGFBI&$   16.2$\tabularnewline
COL23A1&$   15.5$\tabularnewline
SFT2D2&$   14.2$\tabularnewline
\hline
\end{tabular}
}
\caption{Top 10 differentially spliced genes from the models described in \ref{subsubsec:DS}. Tables (a) and (b) show genes with minumum $p$-values, table (c) shows genes with their top scores. Genes highlighted in gray appear in Figures \ref{fig:ANOSVAprobe}, \ref{fig:ANOSVAprobeset}, \ref{fig:FIRMA} and \ref{fig:ZAK}.} 
\end{table}

\subsubsection{Differential gene expression}

In this section we analyse differential gene expression using probe level data. We study two types of transcript clusters: 1) the ones not included in the ANOSVA probe analysis above (with only one probeset or not present in both normal and cancer groups) and 2) the ones not showing differential splicing (ANOSVA probe $p$-value above 0.1). The analysis of group 2) is based on the hierarchical principle: only look for significant main effects (differential expression in this case) among those transcript clusters with no significant interaction terms (differential splicing) \cite[p. 427]{Ekstroem2010}. In total, we analysed 16231 transcript clusters. We fit the following linear model to those transcript clusters:
\begin{equation}
y_{petr} = \mu + \alpha_e + \beta_t + I_r + C_{tr} + \varepsilon_{petr},\label{eq:probeDE}
\end{equation}
where $\beta_t$ is the gene level treatment effect, $\alpha_e$ is a parameter that captures probesets expressed above or below the overall transcript cluster level and $I_r$ and $C_{tr}$ are random effects for patient and chip, respectively. 

\begin{Verbatim}[fontfamily=courier, fontshape=it, fontsize = \relsize{-1.35}]
> aov <- 
+    aov(intensity ~ probeset + treatment + 
+            Error(replicate + replicate:treatment))
> p.F <- 
+   summary(aov)$"Error: Within"[[1]]["treatment", 
+                                        "Pr(>F)"]
\end{Verbatim}

The null hypothesis is that the gene expression is the same in all groups. The top 10 genes, with adjusted $p$-values appear in Table \ref{tab:DE}. The method used for adjusting the $p$-values was Benjamini-Hochberg's correction \cite{Benjamini1995} using the function \texttt{p.adjust(..., method = "BH")}. Only 80 out of 159 genes appearing in Gardina's list of genes up- and down-regulated in tumour \cite[Additional file 1]{Gardina2006} passed our filters and were analysed for differential gene expression. Among those analysed, the genes \textit{CLDN1, SST, MUSK, KIAA1199} and \textit{SLC30A10} were in the top 100. The gene \textit{BEST4} is shown in Figure \ref{fig:DEprobe}. This gene is down-regulated in tumour (blue) compared to normal (red) samples. The thicker lines represent the mean expression levels in the two groups.

\begin{table}[htb]
\begin{center}
\begin{tabular}{|l|r|}
\hline
Gene symbol & $p$-value\tabularnewline
\hline\hline
LARGE&$0.00287$\tabularnewline
IL6R&$0.00847$\tabularnewline
RXRG&$0.00847$\tabularnewline
BAI3&$0.00968$\tabularnewline
C1orf175&$0.01090$\tabularnewline
GRIK3&$0.01090$\tabularnewline
\rowcolor{gray} BEST4&$0.01090$\tabularnewline
KCNN3&$0.01090$\tabularnewline
PLEKHH2&$0.01090$\tabularnewline
HOXD10&$0.01090$\tabularnewline
\hline
\end{tabular}
\end{center}
\caption{Top 10 differentially expressed genes from \eqref{eq:probeDE} with corrected $p$-values. Gene \textit{BEST4}, highlighted in gray, appears in Figure \ref{fig:DEprobe}. \label{tab:DE}} 
\end{table}

\subsection{Gene annotation and visualization} \label{subsec:annotationAndVisualization}
We chose to annotate transcript clusters to genes using the NetAffx transcript cluster annotation release 32 specified in Section \ref{subsec:folderAndFiles} using the \texttt{AnnotateGenes()} function. Some transcript clusters present unspecific annotation and have several possible associated gene names. We have decided to remove such clusters from our output as we cannot map them uniquely to a gene and afterwards interpret the result according to the gene structure.

After gene annotation, the user can select genes for visual inspection.  Visual inspection of candidates for differential splicing is recommended by Affymetrix as a way to identify possible false positives \cite{AffymetrixAlternativeSplicing}. Plots of gene profiles with integrated genomic information are obtained using the \texttt{biomaRt} \cite{biomaRt} and \texttt{GenomeGraphs} \cite{GenomeGraphs} packages in Bioconductor.

We use \texttt{bioMart} to connect to the latest version of the Homo Sapiens dataset in Ensembl \cite{ensembl} (Ensembl genes 68, GRCh37.p8 at the time of writing) and retrieve genes by their HGNC symbol \cite{HGNC}. Gene and exon structures are imported from Ensembl. \texttt{Gene} and \texttt{Exon} objects are created by \texttt{makeGene()} and \texttt{makeTranscript()} from \texttt{GenomeGraphs}. We store expression data and probeset start and stop positions in an \texttt{ExonArray} object by \texttt{makeExonArray()}. The final plot is created passing a list with the objects created to the \texttt{gdPlot(list(exon, gene, transcript,...))} function. 

\begin{figure*}[htp] 
\centering
\includegraphics[width=1\textwidth]{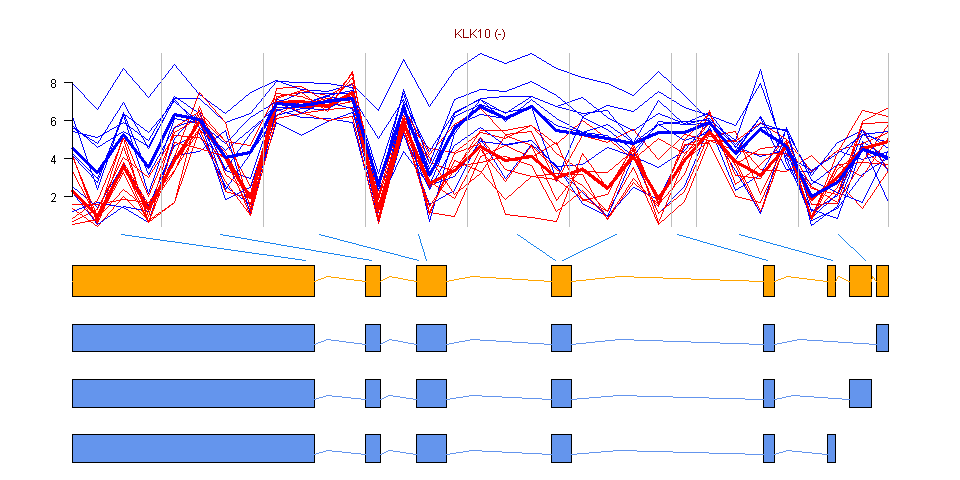}
		\caption{\label{fig:ANOSVAprobe}Profile plot of gene \textit{KLK10} with the gene model and possible transcripts retrieved from Ensembl \cite{ensembl}. The (-) next to the gene name indicates that it is on the reverse strand. The mean intensities of each group are plotted with a thicker line. Exons 5 (mapped by probesets 4 and 5) and 8 (mapped by probeset 9), counted from the 5' end, seem to be higher expressed in tumour samples (blue) than in normal (red).}
\end{figure*}

\begin{figure*}[htp] 
\centering
\includegraphics[width=1\textwidth]{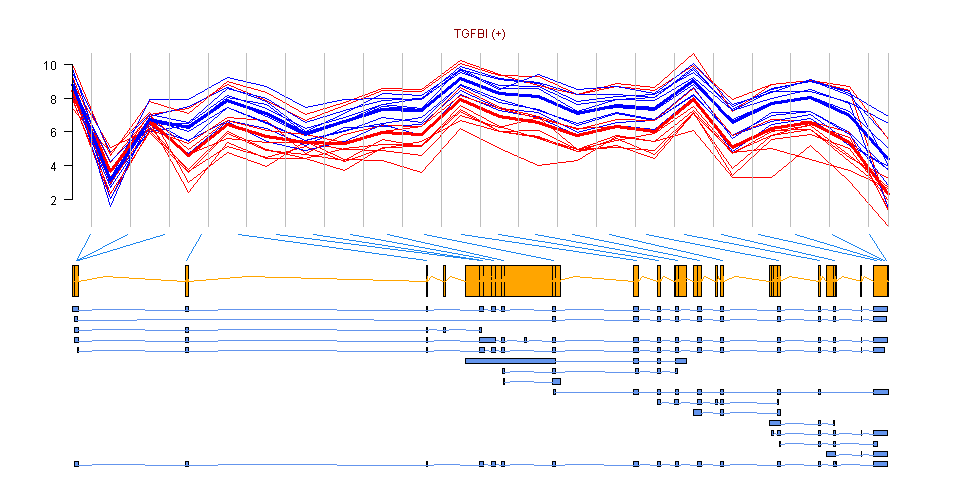}
		\caption{\label{fig:ANOSVAprobeset}Profile plot of \textit{TGFBI} with the gene model and transcripts retrieved from Ensembl \cite{ensembl}. The gene is on the forward strand as indicated by (+) next to its name. The mean intensities of each group are plotted with a thicker line, note that only one estimate is plotted by probeset, it corresponds to the estimate computed by \texttt{ExonRmaPlm(..., mergeGroups=FALSE)}. Here, it seems like the tumour samples (blue) present increased expression from exon 3 until the end of the transcript, with respect to normal (red) samples. Given that \textit{TGFBI} is on the forward strand and that the difference is at the beginning of the transcript, we might be observing a case of alternative promoter usage.}
\end{figure*}

\begin{figure*}[htp] 
\centering
\includegraphics[width=1\textwidth]{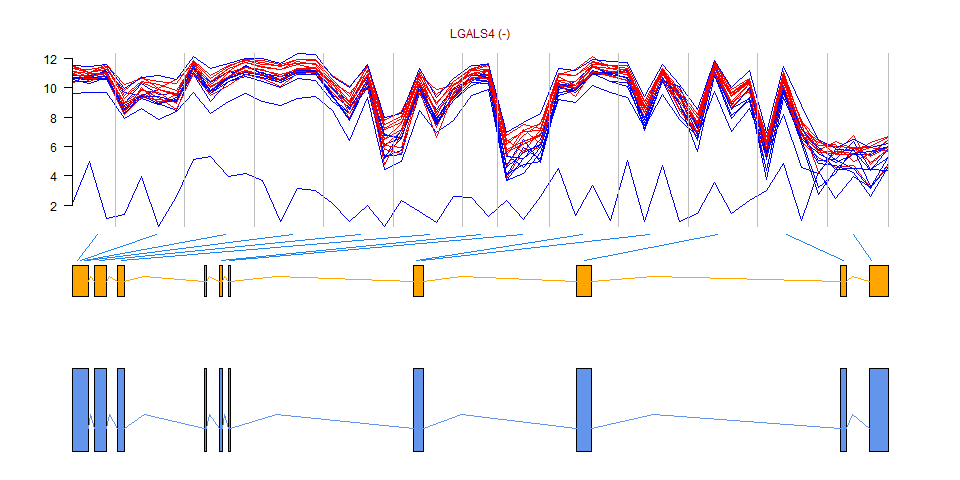}
		\caption{\label{fig:FIRMA}Profile plot of gene \textit{LGALS4} with the gene model and transcripts retrieved from Ensembl \cite{ensembl}, the gene is on the reverse strand. Interestingly, this gene only has one transcript. We might then be facing a novel splicing event or, more likely, a false positive detected by the method.}
\end{figure*}

\begin{figure*}[htp] 
\centering
\includegraphics[width=1\textwidth]{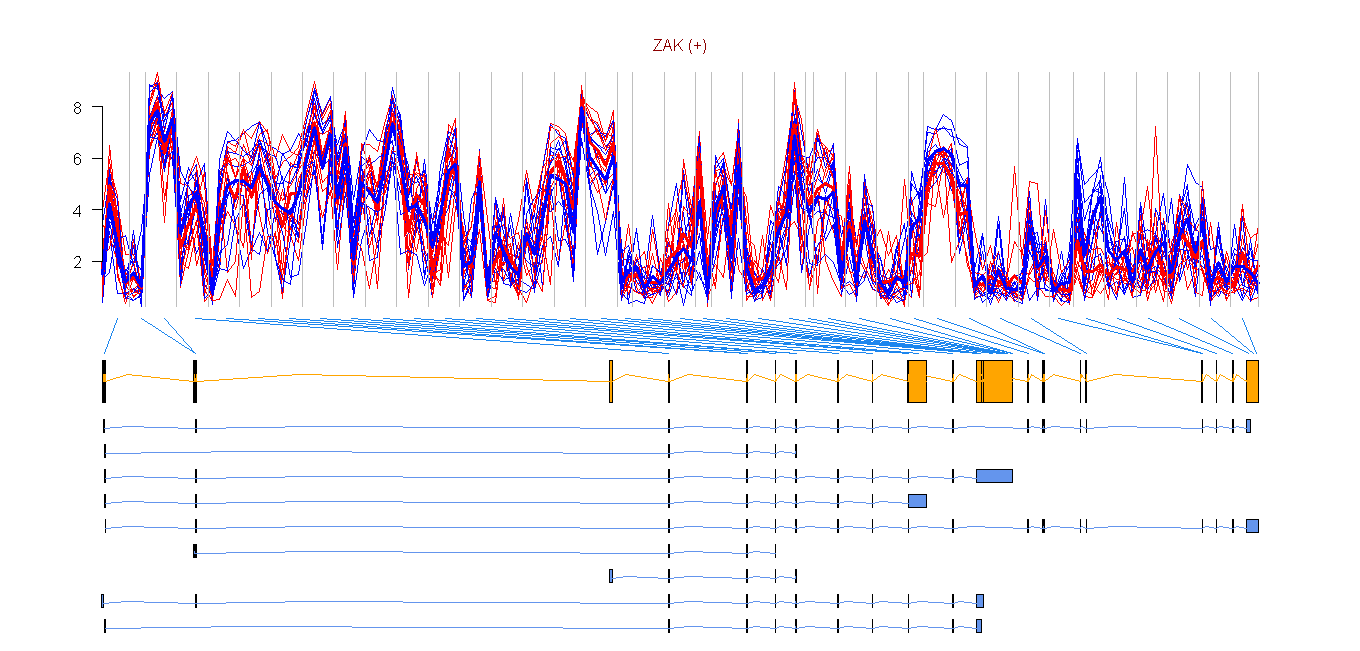}
		\caption{\label{fig:ZAK}Profile plot of gene \textit{ZAK} with the gene model and transcripts retrieved from Ensembl \cite{ensembl}. There seems to be a differential splicing event identified by probesets 5 and 6 (counted from the 3' end of the gene), corresponding to exons 3 and 4 from the 3' end.}
\end{figure*}

\begin{figure*}[htp] 
\centering
\includegraphics[width=1\textwidth]{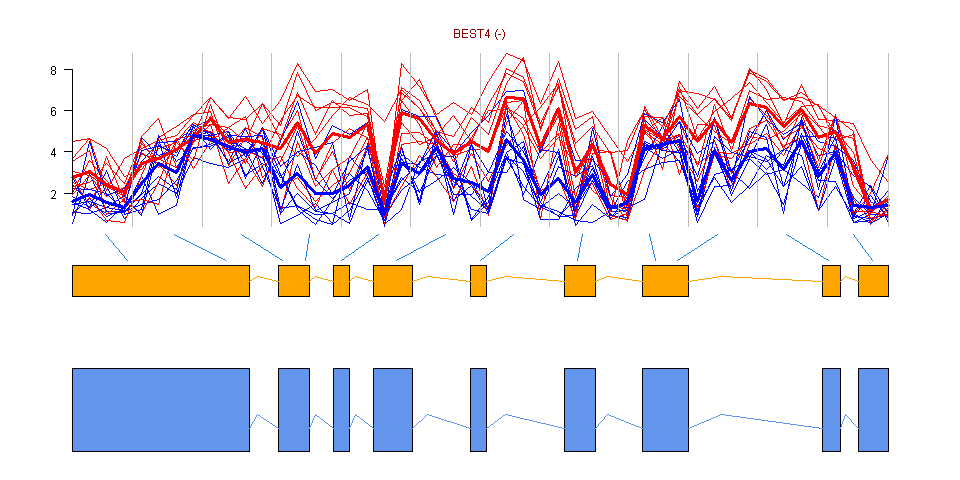}
		\caption{\label{fig:DEprobe}Profile plot of gene \textit{BEST4}, which is on the reverse strand, with the gene model and transcripts retrieved from Ensembl \cite{ensembl}. The gene is under expressed in tumour (blue) compared to normal (red) samples. The thin lines are the expression levels for each sample, while the thicker lines represent the mean intensities in each of the two groups.}
\end{figure*}

Our plots show on top the gene HGNC symbol followed by (+) for genes on the forward strand and by (-) for those on the reverse strand. Below, the plot of probeset intensities appears with vertical lines delimiting probesets. Note that for models based on probe-level data (ANOSVA probe, FIRMA and differential expression), the intensities of all probes in the probeset (1 to 4) are shown. Samples from the same treatment group appear in the same colour, red for normal samples and in blue for tumour samples in this case. Immediately after the profile plot, the gene model retrieved from Ensembl is shown in orange, followed by the possible transcript model(s) in blue. The gene model consists of the exons that appear in all possible transcript models. Exons (boxes) in the gene  model are linked by blue lines to the probesets above, indicating which probeset(s) interrogate which exon. See Figures \ref{fig:ANOSVAprobe} to \ref{fig:DEprobe}.

\section{Discussion}

The aim of this paper was to give a tutorial on how to perform a complete and reproducible analysis of exon array data in \texttt{R}/Bioconductor. We have worked with three packages: \texttt{aroma.affymetrix}, \texttt{biomaRt} and \texttt{GenomeGraphs} to go from .CEL files to intensity data, statistical analysis, annotation and visualization. The packages were chosen for their flexibility and easy of integration. We believe that our workflow covers a number of analysis variants for the exon array, including differential splicing analysis at probe and probeset-level and differential expression analysis at probe level, and gives the user the opportunity to focus on all or only some of the aspects of the data analysis. We make our entire code available so that other researchers can use it as it is or adapt it to their needs.

Different Bioconductor packages could have been used in some of the analysis steps. For example, \texttt{xmapcore} \cite{xmapcore} provides annotation data and cross-mappings between genetic features as transcript clusters or exons and Affymetrix probesets. This package, however, requires the separate installation of a MySQL database, which makes this a more complex alternative than the one we have chosen. The \texttt{xps} package \cite{xps} could have been used for data pre-processing and summarization, but it requires the installation of the \texttt{ROOT} framework \cite{ROOT}, and a certain level of understanding of ROOT files and ROOT trees is recommended. Our workflow does not require any prior knowledge beyond \texttt{R}/Bioconductor. Other free software includes BRB-Array Tools \cite{Simon2007}, based on \texttt{R}, C, Fortran and Java, with an Excel front end, and dChip \cite{Amin2011}, which is written in Visual C{}\verb!++! and developed for Windows - though some users have been able to run it on Mac and Unix computers.

A previous paper on exon arrays \cite{Lockstone2011} suggests a pragmatic approach and does the analysis piecewise starting with Affymetrix Power Tools (APT) and then exporting the data to \texttt{R}. We recognise this is a fix for the lack of straightforward packages for dealing with the exon array in Bioconductor. However, it implies working with several pieces of software so we do not find it fit for reproducible research. Licensed software, like Partek \cite{Partek} or GeneSpring GX \cite{GeneSpring}, has been used in other studies \cite{Zhang2008, Thorsen2011}. In contrast to licensed software, \texttt{R}/Bioconductor is free and available for anyone, it allows the user to control most analysis options and it enables customizable and reproducible analyses that are more easily reviewed. Still, the \texttt{aroma.affymetrix} package does not provide the speed of Affymetrix Power Tools or the licensed software, and it requires more user input. Nevertheless, with this code and minimal user input, any dataset can be analysed regarding differential expression at probe level and differential splicing using the ANOSVA model. 

Differential gene expression could have been analysed using the gene level estimates obtained from the \texttt{plmTr} object in other \texttt{R} packages such as \texttt{limma} \cite{LIMMA}. Another extension of the workflow could include a general analysis strategy of the FIRMA scores, which in this study was tailor-made for a two-treatment scenario. Besides, the profile plots we obtained with \texttt{GenomeGraphs}, though highly informative, are difficult to interpret for genes spanning over a long genomic region, as for example \textit{TGFBI} in Figure \ref{fig:ANOSVAprobeset} and \textit{ZAK} in Figure \ref{fig:ZAK}. In the future, it would be interesting to study the flexibility of the output imported from Ensembl and, for example, remove the intronic regions from the gene and transcript models.

\subsection*{Acknowledgements/Author declaration}
The authors would like to thank Henrik Bengtsson for helpful answers and comments regarding \texttt{aroma.affymetrix}. The authors would also like to thank Hang Phan and Anders Ellern Bilgrau for helpful discussions and ideas about code implementation and data plotting.

MRD designed and implemented the workflow, analysed data and drafted the manuscript. RW assisted in paper drafting and statistical modelling and interpretation. JSB, SF and MKK tested the workflow. JSB, MKK, HEJ and KDS assisted with data interpretation and participated in paper drafting. MB conceived and tested the workflow, assisted in manuscript drafting and coordinated the work. All authors read and approved the final manuscript.

\end{document}